\documentclass[conference,10pt,twocolumn]{IEEEtran}
\usepackage{graphicx}
\usepackage{multibib}
\usepackage{multirow}
\usepackage{epstopdf}
\usepackage{amsmath,amssymb,amsbsy}
\usepackage{subfigure,url,float,cite}
\usepackage[T1]{fontenc}
\usepackage[utf8]{inputenc}

\usepackage[ruled,lined,linesnumbered,noend]{algorithm2e}
\usepackage{comment}

\usepackage{enumitem}
\usepackage{color}

\usepackage[font=small,skip=0pt]{caption}
\usepackage{flushend}
\IEEEoverridecommandlockouts

\begin{document}
\title{{\huge AutoTiering: Automatic Data Placement Manager in Multi-Tier All-Flash Datacenter\thanks{This work was completed during Zhengyu Yang and Morteza Hoseinzadeh's internship at Samsung Semiconductor Inc. This project is partially supported by NSF grant CNS-1452751.}}}


\author{
	\IEEEauthorblockN{
		{Zhengyu Yang}\IEEEauthorrefmark{1},	
		{Morteza Hoseinzadeh}\IEEEauthorrefmark{3},	
		{Allen Andrews}\IEEEauthorrefmark{2},
		{Clay Mayers}\IEEEauthorrefmark{2},\\
		{David (Thomas) Evans}\IEEEauthorrefmark{2},
		{Rory (Thomas) Bolt}\IEEEauthorrefmark{2},	
		{Janki Bhimani}\IEEEauthorrefmark{1},
		{Ningfang Mi}\IEEEauthorrefmark{1}
		and {Steven Swanson}\IEEEauthorrefmark{3}
		}		  	
	
	\vspace{0.1in}
	\IEEEauthorblockA{\IEEEauthorrefmark{1}Dept. of Electrical and Computer
		Engineering, Northeastern University, Boston, MA 02115 
	}

	\IEEEauthorblockA{\IEEEauthorrefmark{3}
	 Dept. of Computer Science and Engineering, University of California San Diego, San Diego, CA 92093
	}	
	
	\IEEEauthorblockA{\IEEEauthorrefmark{2}
	Samsung Semiconductor Inc., Memory Solution Research Lab, Software Group, San Diego, CA 92121
	}
\vspace{-0.5in}
}

\maketitle

\begin{abstract}
In the year of 2017, the capital expenditure of Flash-based Solid State Drivers (SSDs) keeps declining and the storage capacity of SSDs keeps increasing. 
As a result, the ``selling point'' of traditional spinning Hard Disk Drives (HDDs) as a backend storage -- low cost and large capacity -- is no longer  unique, and eventually they will be replaced by low-end SSDs which have large capacity but perform orders of magnitude better than HDDs.
Thus, it is widely believed that all-flash multi-tier storage systems will be adopted in the enterprise datacenters in the near future.
However, existing caching or tiering solutions for SSD-HDD hybrid storage systems are not suitable for all-flash storage systems.
This is because that all-flash storage systems do not have a large speed difference (e.g., 10x) among each tier. 
Instead, different specialties (such as high performance, high capacity, etc.) of each tier should be taken into consideration. 
Motivated by this, we develop an automatic data placement manager called ``AutoTiering'' to handle virtual machine disk files (VMDK) allocation and migration in an all-flash multi-tier datacenter to best utilize the storage resource, optimize the performance, and reduce the migration overhead. 
AutoTiering is based on an optimization framework, whose core technique is to predict VM's performance change on different tiers with different specialties without conducting real migration.
As far as we know, AutoTiering is the first optimization solution designed for all-flash multi-tier datacenters.
We implement AutoTiering on VMware ESXi~\cite{esxi}, and experimental results show that it can significantly improve the I/O performance compared to existing solutions.
\end{abstract}
\vspace{-0.5mm}

\begin{IEEEkeywords}
All-Flash Datacenter Storage, Caching \& Tiering Algorithm, NVMe SSD, Big Data, Cloud Computing, Resource Management, I/O Workload Evaluation \& Prediction
\end{IEEEkeywords}

\vspace{-2mm}
\section{Introduction}
\label{SEC:IN}
\vspace{-2mm}

A basic credendum of cloud computing can be summarized as: user devices are {\em light} terminals to assign jobs and gather results, while all {\em heavy} computations are conducted on remote distributed server clusters.
This {\em light-terminal-heavy-server} structure makes high availability no longer an option, but a requirement in today's datacenters.
Furthermore, when we bring compute, network, and storage capabilities into balance, it is found that the biggest challenge here is closing the gap between compute and storage performance to shift storage's curve back towards Moore's law~\cite{theis2017end}.
In other words, storage I/O is the biggest bottleneck in large scale datacenters.
As shown in study~\cite{andersen2010rethinking},  
the time consumed to wait for I/Os is the main cause of idling and wasting CPU resources, since lots of popular cloud applications are I/O intensive, such as video streaming, file sync and backup, data iteration for machine learning, etc. 
%

To solve the problem caused by I/O bottlenecks, parallel I/O to multiple HDDs in Redundant Array of Independent Disks (RAID) becomes a common approach.
However, the performance improvement from RAID is still limited, therefore, lots of big data applications strive to store intermediate data to memory as much as possible such as Apache Spark.
Unfortunately, memory is too expensive, and its capacity is very limited (e.g., 64$\sim$128GB per server), so it alone is not able to support super-scale cloud computing use cases.
As a device between RAM and HDD, SSD has an acceptable price and space. Since 2008, SSDs started to be adopted in the server market and broke the asymmetric R/W IOPS barrier dramatically, and became one of the promising solutions to speedup storage systems as a cache or a fast tier for slow HDDs~\cite{historySSD}.
%
However, as time goes by, SSD-HDD based solutions are no longer competent to meet the current big data requirements, due to the huge I/O speed gap between SSDs and HDDs.
%
On the other hand, the capital expenditure of Flash-based SSDs keeps decreasing and the storage capacity of SSDs keeps increasing.
Thus, it is widely believed that SSD-HDD solution is just for a transition period, and all-flash multi-tier storage systems will be adopted in the enterprise datacenter in the near future, similar to what happened to HDD-tape solution 30 years ago.
For example, high end NVMe SSDs can replace SATA SSD, and low end TLC SSDs will replace HDDs. 

However, traditional caching algorithms are deemed useful only when the performance difference between storage devices is at least 10x, while the gap between SSD tiers in all-flash datacenters is not that big. 
More importantly, SSDs are expensive, and it is costly to maintain duplicated copies (one for cache and one for backend data) in two SSD tiers.
Thus, we develop an automatic data placement manager called ``AutoTiering'' to handle VM allocation and migration in an all-flash multi-tier datacenter. 
The ultimate goal is to best utilize the storage resource, optimize the performance, and reduce the migration overhead, by associating VMs with an appropriate tier of storage.
AutoTiering is based on an optimization framework to provide the best {\em global} (i.e., for all VMs in the datacenter) migration and allocation solution over runtime. 
AutoTiering's approximation approach further solves the simplified problem in a polynomial time, which considers both historical and predicted performance factors, as well as estimated migration cost.
This comprehensive methodology prevents to frequently migrate back and forth VMs between tiers due to I/O spikes.
Specifically, AutoTiering uses a micro-benchmark-based sensitivity calibration and regression session to predict VM's performance change on different tiers without performing real migration, since different VMs may have different benefits of being upgraded to a high end tier.
We implement AutoTiering on VMware ESXi~\cite{esxi} and evaluate its performance with a set of representative applications.
The experimental results show that AutoTiering can significantly improve the I/O performance by increasing I/O throughput and bandwidth as well as reducing I/O latency.

The rest of this paper is organized as follows. 
Sec.~\ref{SEC:RW} presents the background and literature review.
Sec.~\ref{SEC:FM} formulates the problem and introduces an optimization framework.
Sec.~\ref{SEC:ALG} proposes our approximation algorithm to solve the problem in a polynomial time. 
Experimental evaluation results and analysis are discussed in Sec.~\ref{SEC:EXP}.
We present the conclusions in Sec.~\ref{SEC:CON}.

\vspace{-2mm}
\section{Background and Literature Review}
\label{SEC:RW}
\vspace{-1mm}

Substantial work has been done to improve I/O operations in datacenters at both hardware and software levels. In this section, we discuss some of these work as well as the evolutionary inclination towards AutoTiering. 

%
{\em{SSD as Cache/Tier of HDD:}}
In the recent years, Flash-based SSDs have been commonly used in datacenters. An SSD can be used whether as a cache for HDD or as a distinct storage tier. 
The main difference is that the cache approach has two copies for hot data, one in SSD and one in HDD (two copies are synced under the {\em write through} policy, and are not synced under the {\em write back} policy), while the tiering approach simply migrates data between tiers and only keeps one version of the dataset.
Lots of caching and tiering mechanisms~\cite{LRU-K,SelfCorrectLRU,JT-2Q,YYZ-MQR,LRFU,CLOCK,hoseinzadeh2014reducing,vcs,krish2014hats} are developed for cloud storage systems.

{\em{Data Placement in Multi-tier All Flash Data Center:}}
SSD-HDD based solutions may work for a limited number of users (VMs) with mediate I/O intensity and small working set size, but for the era of super-scale clusters (e.g., clouding computing, IoT in 5G network), the I/O bottleneck gets mitigated, but not resolved~\cite{SieveStore}.
The main reason is that in both SSD-HDD caching and tiering approaches, there still exists a huge performance gap between SSDs and HDDs. 
With the decreasing price and increasing capacity of SSDs, a promising solution to quench this gap is to setup an all-flash datacenter which is becoming a reasonable solution in the near future.
With the aim of further reducing the overall capacity, studies~\cite{xu2014iaware,gmach2009resource} proposed to periodically recompute VM assignments. 
Study~\cite{setzer2012virtual} introduced mathematical model formulations for big data application performance and migrating VMs among tiers, with the aim of minimizing the overhead of data migration. 
Study~\cite{fan2014h} proposed a non-volatile memory based cache policy for SSDs by splitting the cache into four partitions and determining them to their desired sizes according to a page's status.
%
%
We summarize the specs of SSDs with different ends available in market by July 2017 in Table.~\ref{TAB:SSD}. 
As we can see, an all-flash multi-tiers solution can be built from different SSDs with different specialties, e.g., {\em super performance tier} with 3D~XPoint SSD, {\em high performance tier} with NVMe and SLC SSDs, and {\em large capacity tier} with MLC and TLC SSDs.
%
%

\small
\begin{table}[ht]
\centering
\caption{Performance and cost of different SSDs in July 2017.}
\label{TAB:SSD}
\begin{tabular}{|p{1.5cm}|p{1.2cm}|p{1.3cm}|p{1.4cm}|p{1.4cm}|}
\hline
\bf{SSD Type}& {\bf{Cost}} ($\$/GB$)    & {\bf{Max Size}} ($Bytes$)    & {\bf{Read Time}} ($\mu s$)  & {\bf{Write Time}} ($\mu s$)           \\ \hline
3D~XPoint & 4.50   & 375G & 10 &  10\\ \hline
NVMe      & 0.57    &   3.2T        & 20     &  20\\ \hline
SLC SATA3      & 0.64   & 480G     & 25    &  250\\ \hline
MLC SATA3 & 0.30   & 2T      & 50    &  750\\ \hline
TLC SATA3      & 0.28      & 3.84T       & 75     &  1125\\ \hline
\end{tabular}
\end{table}
\normalsize

\vspace{-2mm} 
\section{Problem Formulation}
\label{SEC:FM}
\vspace{-1mm}

In this section, we first formulate the problem of VM allocation and migration in an all-flash multi-tier datacenter, and then develop an optimization framework to best utilize SSD resource among VMs in the datacenter.

\vspace{-1mm} 
\subsection{System Architecture}
\label{system-architecture}

\begin{figure}[t]
\centering
\includegraphics[width=0.45\textwidth]{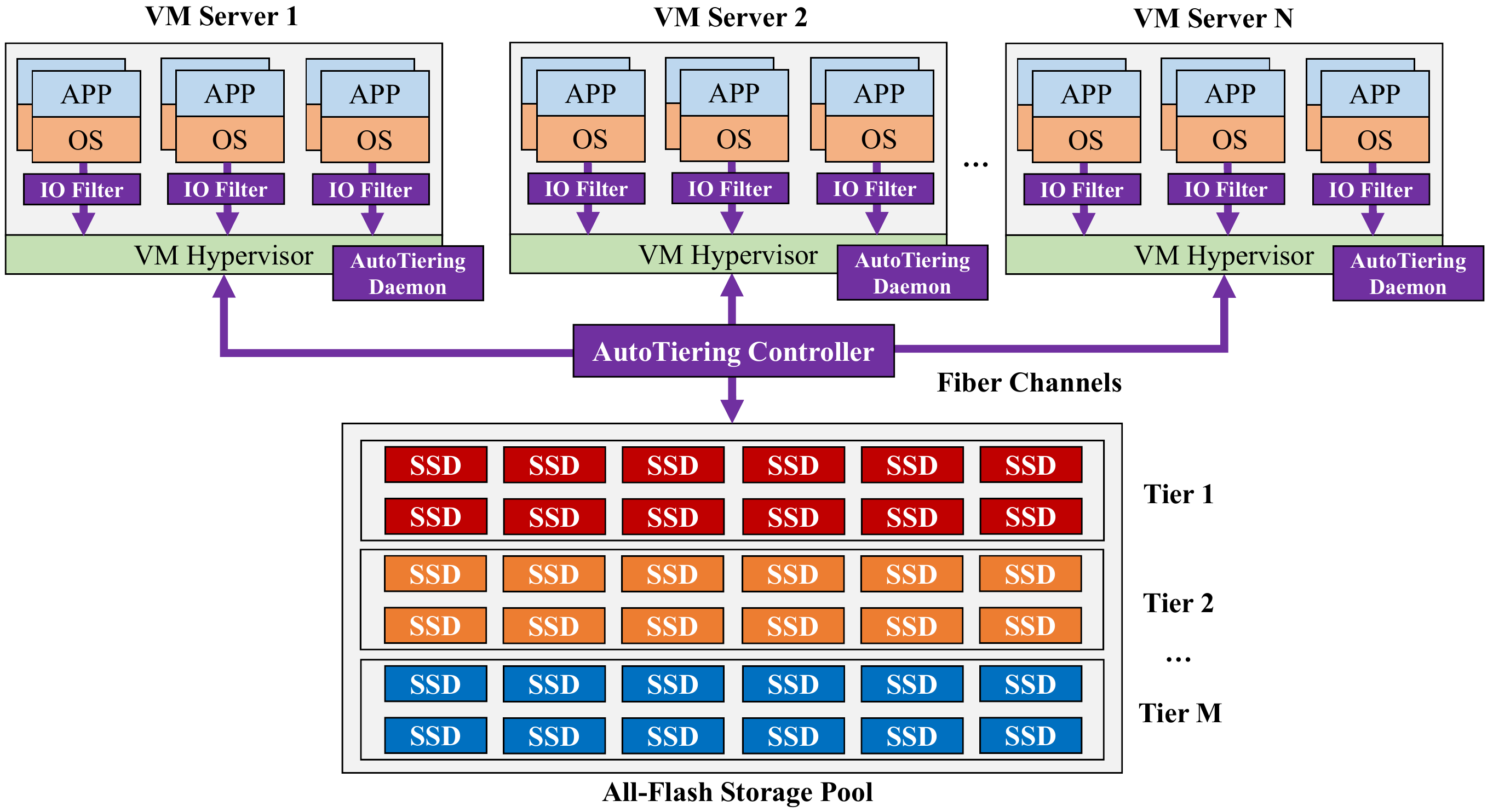}
\caption{\small Architecture of AutoTiering in a multi-tier all-flash storage system.}
\label{FIG:ARCH}
\end{figure}

Fig.~\ref{FIG:ARCH} illustrates the system architecture of AutoTiering which has the following components:
\begin{itemize}
    %
    \item{{\em{IO Filter:}} Attached to each VMDK (virtual machine disk file) being managed on each host, it is responsible for collecting I/O related statistics as well as running special latency tests on every VMDK. The data will be collected and sent to the AutoTiering Daemon on the host~\cite{FILTER}.} 
    \item{{\em{Daemon}}: Running on the VM hypervisor of all hosts, it tracks the workload change (i.e., I/O access pattern change) of each VM, collects the results of latency injection tests from the IO Filter, and sends them to the Controller.}
    \item{{\em{Controller:}} Running on a dedicated server, it is responsible for making decisions to trigger migration based on the predicted VM performance (if it is migrated to other tiers) and the corresponding migration overhead.}
\end{itemize}


\vspace{-1mm} 
\subsection{Optimization Framework}
\label{optimization-framework}

To develop an optimization framework aimed at minimizing the total amount of server hours by determining a VM migration schedule, we formulate the problem by investigating the following factors: 
First, from each VM's point of view, the reason for a certain VM to be migrated from one tier to another is that the VM can perform better (e.g., less average I/O latency, higher IOPS, higher throughput, etc.) after migration.
Second, the corresponding migration cost need to be considered, because migration is relatively expensive (consumes resource and time) and not  negligible.
Third, from the {\em global optimization}'s point of view, it is hard to satisfy all VMs to be migrated to their favorite tiers at the same time due to resource constraints and their corresponding SLAs (i.e., Service Level Agreement).
Fourth, the {\em global optimization} should consider overtime changes of all VMs as well as post-effects of migration. For example, the current best allocation solution may lead to a bad situation for the future since VMs are changing behaviors during runtime.
%
%
Based on these factors, our optimization framework needs to consider potential benefits and penalties, migration overhead, historical and predicted performance of VMs on each tier, SLA, as shown in  Eq.~\ref{EQ:OBJ} to~\ref{EQ:CONS}. 
Table~\ref{TAB:NOTE} represents notations that we use in this paper.

\small
\begin{alignat}{2}
&\textbf{Maximize: }   \nonumber \\   
& \sum _{ \forall v,\forall t,\forall \tau  }^{  }{ w_{ v,\tau  }\cdot[\sum _{ \forall k }^{  }{ \alpha _{ k }\cdot r(v,t_{ v,\tau  },\tau ,k) } -\beta \cdot g(v,t_{ v,\tau -1 },t_{ v,\tau  })] }  ,  \label{EQ:OBJ} \\ 
&\textbf{Subject to: } \nonumber\\
& \text{for  }  \forall v,\forall \tau :\nonumber\\
& \quad t_{ v,\tau  }\neq \emptyset, \text{and } | t_{ v,\tau  }| \geq 1,   \label{EQ:MUSTASSIGN} \\
& \text{for  }  \forall \tau _{ 1 },  \tau _{ 2 }\in [0,+\infty ): \nonumber\\
& \quad r(v,t_{ v,\tau _{ 1 } },\tau _{ 1 },k_{ s })\equiv r(v,t_{ v,\tau _{ 2 } },\tau _{ 2 },k_{ s }) \geq 0,  \label{EQ:SIZE} \\
& \text{for  }  \forall v,\forall t,\forall \tau: \nonumber\\
& \quad r(v,t_{ v,\tau  },\tau ,k)={ { r }_{ { Prd } } }(v,t_{ v,\tau -1 },t_{ v,\tau  },\tau -1,{ k }) \geq 0, \label{EQ:PRED} \\
& \quad \sum_{\forall v}^{}{ r(v,t_{ v,\tau  },\tau ,k) \leq \Gamma_k \cdot R(t_{v,\tau},k)},  \label{EQ:CONS}
%
\end{alignat}
\normalsize

\vspace{-2mm}

The main idea is to maximize the ``{\em Profit}'', which is the entire performance gain minus penalty, i.e., ``{\em Performance Gain - Performance Penalty}'', as shown in the objective function Eq.~\ref{EQ:OBJ}. 
The inner ``sum'' operator conducts a weighted sum of the usage of all types of resources (such as throughput, bandwidth and storage size, etc.) of each VM, assuming migrating VM $v$ from tier $t_{\tau-1}$ to $t_{\tau}$. 
The outer ``sum'' operator further iterates all possible migration cases, where weight parameter $w_{v,\tau}$ reflects the SLA of VM $v$ in $\tau$ epoch.
Notice that the term ``migration'' in this paper is not to migrate a VM from one host server to another.
Instead, only backend VMDK files are migrated from one to the other SSD tier. 
As a result, non-disk-I/O related resources (e.g., host-side CPU and memory) are not considered in this paper. 
Eq.~\ref{EQ:MUSTASSIGN} guarantees that each VM is hosted by at least one disk tier. 
In fact, each VM can have multiple VMDKs, and each of them can be located at different tiers.
Unlike the previous SSD-HDD tiering work~\cite{cave,vcs} that are   operating on fine-grained data blocks due to HDD speed bottleneck, the minimal unit to migrate in this paper is the entire VMDK of each VM. 
%
Eq.~\ref{EQ:SIZE} ensures that storage size (i.e., VMDK size) will not change before and after migrations, where $k_s$ is the type index of storage resource.
Eq.~\ref{EQ:PRED} shows that a prediction model function is utilized to predict the performance gain (for details, see Sec.~\ref{tier-speed-sensitivity-calibration}). 
Eq.~\ref{EQ:CONS} lists resource constraints, where $\Gamma_{k}$ is preset upper bound (in percentage) of each type (i.e., $k$-th) of resource  that can be used.
%
Finally, the temporal migration overhead is the size of the VM to be migrated, divided by the bottleneck of migrate-out read speed and migrate-in write speed, i.e.,
$ g(v,t_{ v,\tau -1 },t_{ v,\tau  })= \frac { r(v,t_{ v,\tau -1 },\tau -1,k_{ s }) }{\mu_(v,\tau-1)}$.
The migration speed is $\mu_(v,\tau-1)= min([Pr(\Lambda ,t_{ v,\tau -1 },\tau -1) + Pw(\Lambda ,t_{ v,\tau -1 },\tau -1))$, where 
$Pr(\Lambda ,t_{ v,\tau -1 },\tau -1)$ is the function of available remaining read throughput.
Since we are going to {\em live migrate} VM $v$, the read throughput used by this VM ($Pr(v,t_{ v,\tau -1 },\tau -1)$) is also available and has been added back here. 
$Pw(\Lambda ,t_{ v,\tau -1 },\tau -1)$ gives the migrate-in write throughput.

\small
\begin{table}[t]
\small
\centering
\caption{Notations.}
\label{TAB:NOTE}
\begin{tabular}{|p{2.7cm}|p{5.4cm}|}
\hline
\bf{Notation}      & \bf{Meaning}           \\ \hline
$v$, $v_i$ & VM $v$, $i$-th VM, $v\in[1,v_{max}]$, where $v_{max}$ is the last VM. \\ \hline
$t$, $t_i$ & Tier $t$, $i$-th tier, $t\in[1,t_{max}]$, where $t_{max}$ is the last tier.  \\ \hline
$t_{v,\tau}$ & VM $v$'s hosting tier during epoch $\tau$.  \\ \hline
$k$  & Different types of resources, $k\in[1,k_{max}]$, such as throughput, bandwidth, storage size, and etc.  \\ \hline
$\tau$ & Temporal epoch ID, where $\tau\in[0,+\infty)$. \\ \hline
 $\alpha_k$, $\beta$ & $k$-th resource's weight, migration cost weight. \\ \hline
$r(v,t_{ v,\tau  },\tau ,k)$ & Predicted type of $k$ resource usage of VM $v$ running on tier $t_{v,\tau}$.  \\ \hline
$g(v,t_{ v,\tau -1 },t_{ v,\tau  })$ & Migration cost of VM $v$ during epoch $\tau$ from tier $t_{ v,\tau-1  }$ to tier $t_{ v,\tau  }$. \\ \hline
$k_{s}$ & ``Storage'' resource's index. \\ \hline
$\mu(v,\tau-1)$ & Estimated migration speed of VM $v$ at time $\tau-1$ epoch.\\\hline
$Pr(v,t,\tau)$,  $Pw(v,t,\tau)$ & Read/write resource of VM $v$ on tier $t$ during epoch $\tau$.   \\ \hline
$Pr(\Lambda,t,\tau)$, $Pw(\Lambda,t,\tau)$ & All remaining available read/write resource of tier $t$ during epoch $\tau$. \\ \hline
$\Gamma_{k}$ & Upper bound (in percentage) of each type of resource that can be used. \\ \hline
$R(t_{v,\tau},k)$ & Total capacity of $k$-th type of resource on tier $t_{v,\tau}$. \\ \hline
$L_t$  & Original average I/O latency (without injected latency) of tier $t$.\\ \hline
$b_v$, $m_v$ & Parameters of TSSCS liner regression model ($y=mx+b$).\\ \hline
$s_v$  & Average I/O size of VM $v$.\\ \hline
$S_v$, VM{[}v{]}.size & Storage size of VM $v$.\\ \hline
$w_{v,\tau}$, $wetP[t]$, $wetB[t]$, $wetS[t]$   & Weight of VM $v$ and weights of tier $t$'s each type of resource. \\\hline 
$maxP[t]$, $maxB[t]$, $maxS[t]$, $spc[t].P$, $spc[t].B$, $spc[t].S$ & Preset available resource caps and specialties of tier $t$, P=throughput, B=bandwidth, S=storage size. \\\hline
\end{tabular}
\end{table}
\normalsize

Since the system has no information on the future workload I/O patterns, it is impossible to conduct the {\em global} optimization for all $\tau$ time periods during runtime.
Moreover, the decision making process for each migration epoch in the {\em global} optimal solution depends on the past and future epochs, which means that Eq.~\ref{EQ:OBJ} cannot be solved by traditional sub-optimal-based dynamic programming techniques. 
Lastly, depending on the complexity of the performance prediction model (e.g., Eq.~\ref{EQ:PRED}), the optimization problem can easily become NP-hard.

%

\vspace{-1mm}
\section{Approximation Algorithm Design}
\label{SEC:ALG}
\vspace{-1mm}

To obtain the result close to the optimized solution in a polynomial time,  we have to relax some constraints.
%
In detail, we first downgrade the goal from ``\emph{global optimizing for all time}'' to ``\emph{only optimizing for each epoch}'' (i.e., runtime greedy).
Furthermore, since we have the foreknowledge of each tier's performance ``specialty'' (such as high throughput, high bandwidth,  large space, small write amplification function, large over-provisioning ratio, large program/erase cycles, etc.), we can make migration decisions based on the ranking of the estimated performance of each VM on each tier, considering each tier's specialties and corresponding estimation of migration overhead.
Details of our approximation algorithm are discussed in the following subsections.

\subsubsection{Main Procedure}
\label{main-procedure}

Alg.~\ref{ALG:1} lines 1-8 show the main procedure of AutoTiering, which periodically monitors 
the performance and checks 
whether a VM needs to be migrated.
Specifically, $monitorEpoch$ is the frequency of evaluating and regressing the performance estimation model, and  $migrationEpoch$ is the frequency of triggering VM migration from one tier to the other one.
The migration decision is made at the beginning of each $migrationEpoch$, which is greater than $monitorEpoch$.
Apparently, the smaller temporal window sizes, the more frequent monitoring, measurement, and migration. 
The system administrator can balance a tradeoff between the accuracy and the migration cost by conducting a sensitivity analysis before deployment.
As shown in line 4 of Alg.~\ref{ALG:1}, procedure $tierSpdSenCalibrate$ estimates VM's performance on each tier based on the regression model. 
Line 5, thus, calculates performance matrices, and line 6 calculates the score by considering the historical and current performance of each VM, estimates the corresponding migration overhead. 
Finally, VM migrations are triggered, see line 8.
We describe their details in the following subsections.

%
%

\begin{algorithm}[h]
\small
\SetAlFnt{\footnotesize}
\SetKwProg{ProcPrv}{Procedure}{}{}
\SetKwFunction{FuncPrv}{autoTiering}
\SetKwProg{ProcTier}{Procedure}{}{}
\SetKwFunction{FuncTier}{tierSpdSenCalibration}
\SetKwProg{ProcCap}{Procedure}{}{}
\SetKwFunction{FuncCap}{calCapacityMatrices}
\SetKwProg{ProcLat}{Procedure}{}{}
\SetKwFunction{FuncLat}{estimateAvgLat}
\SetKwIF{If}{ElseIf}{Else}{if}{then}{else if}{else}{endif}
\ProcPrv{\FuncPrv{}}
{
	\While{$True$}
	{
	    \If{$currTime$ MOD $monitorEpoch = 0$}
	    {
	        $tierSpdSenCalibration()$\;
            $calCapacityMatrices()$\;
            $calScore()$\;
	    }
	    \If{$currTime$ MOD $migrationEpoch = 0$}
	    {
            $triggerMigration()$\;  
	    }
	}
}
\ProcTier{\FuncTier{}}
{
	\For{t $\in$ tierSet}
	{
	    \For{v $\in VMSet[t]$}
	    {
	        \For{samplesWithLatency $\in$ sampleSet[t][v]}
	        {
                $CV+=calCV(samplesWithLatency)$\;
                $samplesWithLatency=avg(samplesWithLatency)$\;
	        }
	        $CV/=len(sampleSet[t][v])$\;
            \If{$CV \geq 1$}
            {
                $VM[v].conf = 0.05$\;
            }
            \Else
            {
                $VM[v].conf = 1-CV$\;
            }
            $(VM[v].m,VM[v].b)=regress(sampleSet[t][v])$\;
        }
    }
	\KwRet\;
}
\ProcCap{\FuncCap{}}
{
	\For{t $\in$ tierSet}
	{
	    \For{v $\in VMSet[t]$}
	    {
	    $Lat = estimateAvgLat(v,t)$\;
	    \If{Lat > 0}
	    {
	        $IOPS = 10^6 / Lat$\;
	    }
	    \Else
	    {
	        $IOPS = 0$\;
	    }
        $VMCapMat[t][v].P = IOPS$\;
        $VMCapMat[t][v].B = \frac{IOPS \times VM[v].avgIOSize}{10^6}$\;
        $VMCapMat[t][v].S = VM[v].size$\;
	    }
	}
	\KwRet\;
}
\ProcLat{\FuncLat{v,t}}
{
	\KwRet $VM[v].m \times (tierLatency[t] - tierLatency[VM[v].tier])) + VM[v].b$\;
}
\caption{AutoTiering Procedure Part I.}
\label{ALG:1}
\end{algorithm}

\subsubsection{Tier Speed Sensitivity Calibration}
\label{tier-speed-sensitivity-calibration}

In order to estimate VM's performance on other tiers without conducting actual migration, we first try to ``emulate'' the speed of tiers by manually injecting a synthetic latency to each VM's I/Os, and measure the resultant effect on total I/O latency by calling IOFilter APIs. 
%
%
Our preliminary experiments show that the performance variation can be modeled to a linear function.
VMs running different types of applications have varying performance sensitivity to the tier.
Motivated by these observations, we introduce a micro-benchmark session, called ``Tier Speed Sensitivity Calibration Session ({\texttt{TSSCS}})'', to predict (i.e., without conducting actual migration) how much performance benefit (resp. performance penalty) it can take for each VM if we migrate that VM to a faster (resp. slower) tier.
In detail, {\texttt{TSSCS}} has the following properties:

\begin{itemize}
\item {{\em{Lightweight}}: Running inside the IOFilter, {\texttt{TSSCS}} injects a synthetic tiny latency to each VM in a very low frequency, without affecting the current hosting workload performance.}
\item {{\em{Multi-Samples per Latency}}: {\texttt{TSSCS}} improves the accuracy of emulating each VM's performance under each tier by taking the average over multiple samples that are obtained with the same injected latency of each tier.}
\item {{\em{Multi-Latencies per Session}}: {\texttt{TSSCS}} takes multiple latencies per session to refine the regression. 
}
\item {{\em{Multi-Sessions during Runtime}}: {\texttt{TSSCS}} is periodically triggered to update the regression model by calling
  ``$tierSpdSenCalibration$''  in Alg.~\ref{ALG:1} line 4.}
\end{itemize}

\begin{figure}[t]
\centering
\includegraphics[width=0.4\textwidth]{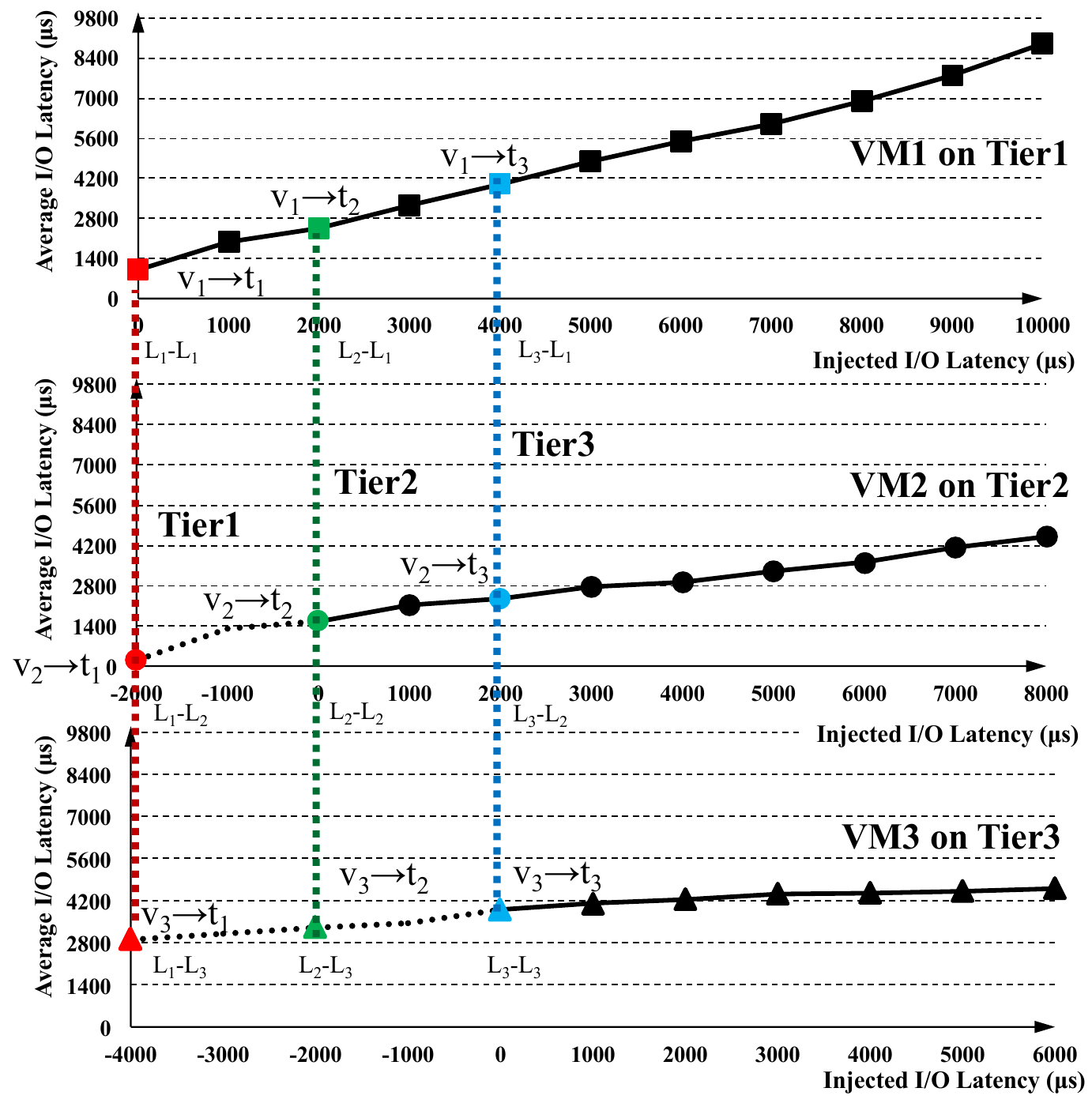}
\caption{\small Example of average I/O latency estimation.}
\label{FIG:CURVE}
\end{figure}

Fig.~\ref{FIG:CURVE} depicts an example of three VMs running on three different tiers: VM $v_1$ on tier $t_1$, VM $v_2$ on tier $t_2$, and VM $v_3$ on tier $t_3$. 
Assume $t_1$ is 2,000~$\mu s$ faster than $t_2$, and $t_2$ is 2,000~$\mu s$ faster faster than $t_3$. 
We run {\texttt{TSSCS}} on each VM on their hosting tier, and get the latency curves shown in three plots in Fig.~\ref{FIG:CURVE}. 
Since all injected latencies are additional to the original bare latency (i.e., without injected latency), we have to align them according to the absolute latency values (i.e., bare latency + injected latency).
Notice that since we cannot inject negative latencies (i.e., obviously we can only slow down the tier), 
the dash lines in subfigures ``VM2 on tier2'' and ``VM3 on tier3'' are regressed based on the solid lines.
After that, we can draw three (colored) lines for each tier based on their absolute latency values (i.e., red for tier1, green for tier2, and blue for tier3). 
Then, we can easily predict the average I/O latency values of each VM on each tier (i.e., red points for $t_1$, green points for $t_2$, and blue points for $t_3$).
%
We see that VM1 is the most sensitive one to tier speed changes (i.e., with the greatest gradient), while VM3 is the least sensitive one (i.e., relatively flat). 
Therefore, intuitively, we should assign VM1 to tier1 (fastest tier) and VM3 to tier3 (slowest tier).
Alg.~\ref{ALG:1} lines 9-21 describe the procedure of tier speed sensitivity calibration session ({\texttt{TSSCS}}). 
In addition, AutoTiering calculates the coefficient variation (CV) of sampling results to decide the estimation confidence, see in lines 13 and 16-19.

\begin{algorithm}[ht]
\small
\SetAlFnt{\footnotesize}
\SetKwProg{ProcPrv}{Procedure}{}{}
\SetKwFunction{FuncPrv}{calScore}
\SetKwProg{ProcPub}{Procedure}{}{}
\SetKwFunction{FuncPub}{migCost}
\SetKwProg{ProcMig}{Procedure}{}{}
\SetKwFunction{FuncMig}{triggerMigration}
\SetKwIF{If}{ElseIf}{Else}{if}{then}{else if}{else}{endif}
\ProcPrv{\FuncPrv{}}
{
	\For{t $\in$ tierSet}
	{
	    \For{v $\in VMSet[t]$}
	    {
	    \If{$maxP[t]<VMCapMat[t][v].P$ or
               $maxB[t]<VMCapMat[t][v].B$ or
               $maxS[t]<VMCapMat[t][v].S$}
            {
               $VMCapRatMat[t][v].P = 0$\;
               $VMCapRatMat[t][v].B = 0$\;
               $VMCapRatMat[t][v].S = 0$\;
               $tierVMPerfScore[t][v] = -1$\;
               continue\;
            }
        /* Convert to percent capacities */\;
            $VMCapRatMat[t][v].P =\frac{VMCapMat[t][v].P}{maxI[t]}$\;
            $VMCapRatMat[t][v].B = \frac{VMCapMat[t][v].B}{maxP[t]}$\;
            $VMCapRatMat[t][v].S = \frac{VMCapMat[t][v].S}{maxS[t]}$\;
            $tierVMPerfScore[t][v]=agingFactor\times ttlVMCapRatMat[t][v] + currCapScore(t,v) - wetMig[t]\times migCost(t,v)$\;
	    }
	    }
	\KwRet\;
}
\ProcPub{\FuncPub{t,v}}
{
    $migSpd=min(remReadThrput(VM[v].tier)+VM[v].currReadThrpt, remWriteThrput(t))$\;
    \KwRet $VM[v].size/migSpd$\;
}
\ProcMig{\FuncMig{}}
{
	\For{t $\in$ tierSet}
	{
	    \For{v $\in descendingSortByScore(tierVMPerfScore[t])$}
	    {
	    \If{$VM[v].isAssigned=False$ and $tierVMPerfScore[t][v] \neq -1$ and $tierHasCapacityForVM(t,v)$}
	    {
	        assignVMToTier(v,t)\;
            VM[v].isAssigned=True\;
	    }
	    }
	}
	\KwRet\;
}
\caption{AutoTiering Procedure Part II.}
\label{ALG:2}
\end{algorithm}

\subsubsection{Performance Capacity Matrices}
\label{performance-matrices}

Once we have the average latency vs. injected latency curves of each VM of the current moment, we calculate corresponding performance estimation of throughput (denoted as $P$, unit in IOPS), bandwidth (denoted as $B$, unit in MBPS), and storage size (denoted as $S$, unit in bytes), and record them into three two-dimensional matrices, i.e., $VMCapMat[t][v].P$, $VMCapMat[t][v].B$, and $VMCapMat[t][v].S$, where $t$ and $v$ are IDs of tier and VM, respectively.
Compared to the first two matrices, the last matrix is relatively straightforward to be obtained by calling the hypervisor APIs to measure the storage size that each VM is occupying.  
%
As shown in Alg.~\ref{ALG:1} lines 22-33, AutoTiering updates the VM capacity matrices (i.e., $VMCapMat$) by reiterating for all tiers and VMs. It estimates ``new'' latency under other tiers by calling the ``$estimateAvgLat(v,t)$'' function in Alg.~\ref{ALG:1} line 25. 
Lines 34-35 show the detail of $estimateAvgLat$ function, where the input parameters are VM $v$ and target tier $t$ for estimation, which returns an estimation based on linear regressions of $m$ and $b$ values.  
Once the estimated average I/O latency results are obtained, 
we calculate the throughput and bandwidth in Alg.~\ref{ALG:1} lines 30 and 31.
Lastly, the storage size will also be updated into the $VMCapMat$ (i.e., Alg.~\ref{ALG:1} line 32). 
%
%
Furthermore, since it gets harder to evaluate demands of different recourse types together (because they have different units), AutoTiering normalizes the VM's estimated/measured throughput, bandwidth and storage utilization value according to the total available resource capacity of each tier, which is called the normalized capacity utilization rate matrix $VMCapRatMat$, as shown in Alg.~\ref{ALG:2} lines 4-13. 

\subsubsection{Performance Score Calculation}
\label{score-calculation}

AutoTiering takes three steps to calculate the performance score based to reflect the following factors:
First, {\em{Characteristics of both tier and VM}}: the score should reflect each tier's specialty and each VM's workload characteristics running on each tier. Thus, our solution is to calculate each VM's score on each tier separately.
Second, {\em{SLA weights}}: VMs are not equal since they have different SLA weights, as shown in Eq.~\ref{EQ:OBJ}.
Third, {\em{Confidence of estimation}}: we use coefficient variation calculated in performance matrices to reflect the confidence of estimation.
Fourth, {\em{History and migration costs}}: a convolutional aging factor for historical scores and estimated migration cost are considered during the score calculation.

{\em{[Step 1] Tier Specialty Matrix}}:
\label{tier-specialty-matrix}
To reflect the specialty, we introduce a two-dimension tier-specialty matrix $spc$.
For example, ``$spc[t].P=1$, $spc[t].B=1$ and $spc[t].S=0$'' reflects that tier $t$ is good at throughput and bandwidth, but bad at storage capacity.
In fact, this matrix can be extended to a finer granularity to further control specialty degree, and  more types of resources can be included into this matrix, if needed. 
Moreover, tiers are sorted in the order of high-to-low-end (e.g., most-to-least-expensive-tier) in the matrix, and this order is regarded as a priority order during the migration decision making period. 

{\em{[Step 2] Orthogonal Match between VM Demands and Tier Specialties:}}
\label{orthogonal-match-between-vm-demands-and-tier-specialties}
The next question is ``{\em {how to reflect each VM's performance on each tier AND reflect how good VMs can utilize each tier's specialty?}}''. 
Our solution is to introduce a process called ``orthogonal match'' (denoted as ``$\Omega$'') to score the ``matchness''.
This process is a per-VM-per-tier multiplication operation of ``specialty'' matrix and `$`VMCapRatMat$'' matrix, i.e.,
\vspace{-3mm}
\small
\begin{alignat}{2}
&currCapScore(t,v)=\Omega (t,v)\nonumber\\
&=\begin{bmatrix} spc[t].P\times wetP[t], &spc[t].B\times wetB[t], &spc[t].S\times wetS[t] \end{bmatrix}  \nonumber\\ 
&\times \begin{bmatrix} VMCapRatMat[t][v].P \\ VMCapRatMat[t][v].B \\ VMCapRatMat[t][v].S \end{bmatrix}
 \times VM[v].SLA \times VM[v].conf \nonumber\\
& \div  (wetP[t]+wetB[t]+wetS[t]) ,
\end{alignat}
\normalsize
where $currCapScore$ gives the current capacity score,  and $VMCapRatMat$ is the VM capacity utilization rate matrix.

{\em{[Step 3] Convolutional Performance Score:}}
\label{convolutional-score}
The final performance score is a convolutional sum of historical score, current epoch capacity score and penalty of corresponding migration cost, i.e., 
\vspace{-2mm}
\small
\begin{alignat}{2}
&tierVMPerfScore = agingFactor \times histTierVMPerfScore \nonumber \\
&+ currCapScore - wetMig \times migCost
\end{alignat}
\normalsize
This process is also shown in Alg.~\ref{ALG:2} line 14.
Specifically, to avoid the case that some VMs are frequently migrated back and forth between tiers (due to making decision only based on recent one epoch which may contain I/O spikes or bursties), AutoTiering needs to convolutionally consider history scores, with a preset $agingFactor$ to fadeout outdated scores. 
Current capacity score $currCapScore$ is calculated by the orthogonal match procedure. 
%
Additionally, Alg.~\ref{ALG:2} lines 16-18 show the procedure of $migrationCost$ calculation. 
\vspace{-2mm}
\section{Performance Evaluation}
\label{SEC:EXP}

\begin{figure*}[ht]
\centering
\includegraphics[width=0.86\textwidth]{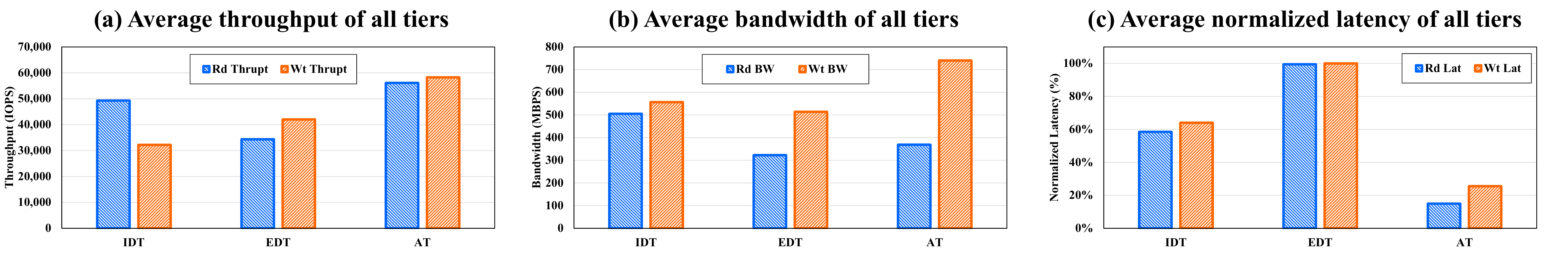}
\caption{\small Average throughput, bandwidth, and latency of all tiers.}
\label{FIG:REC-AVG}
\end{figure*}

\begin{figure}[h]
\centering
\includegraphics[width=0.37\textwidth]{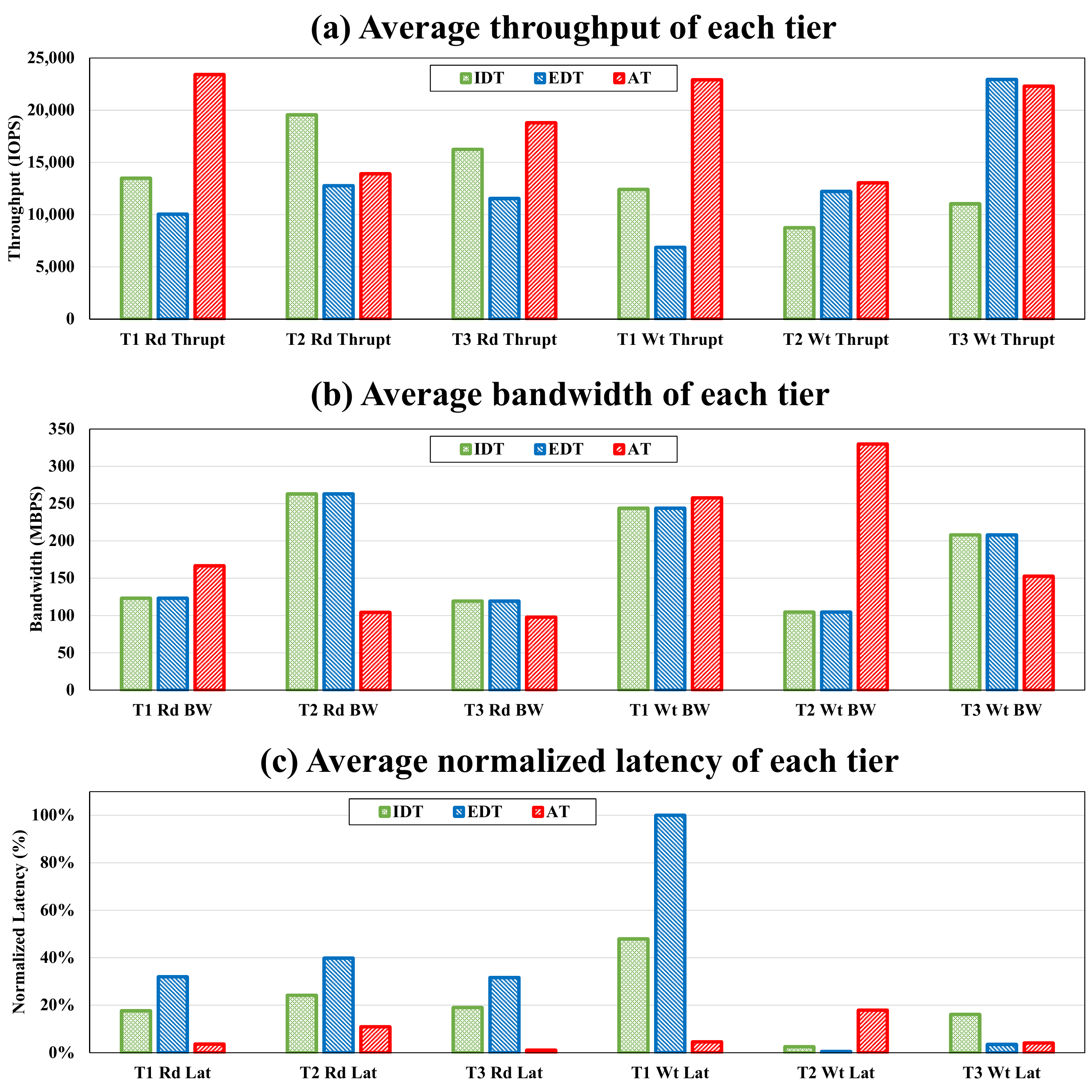}
\caption{\small Average throughput, bandwidth, and latency of each tier.}
\label{FIG:REC-AVGTIER}
\end{figure}

\begin{figure}[h]
\centering
\includegraphics[width=0.37\textwidth]{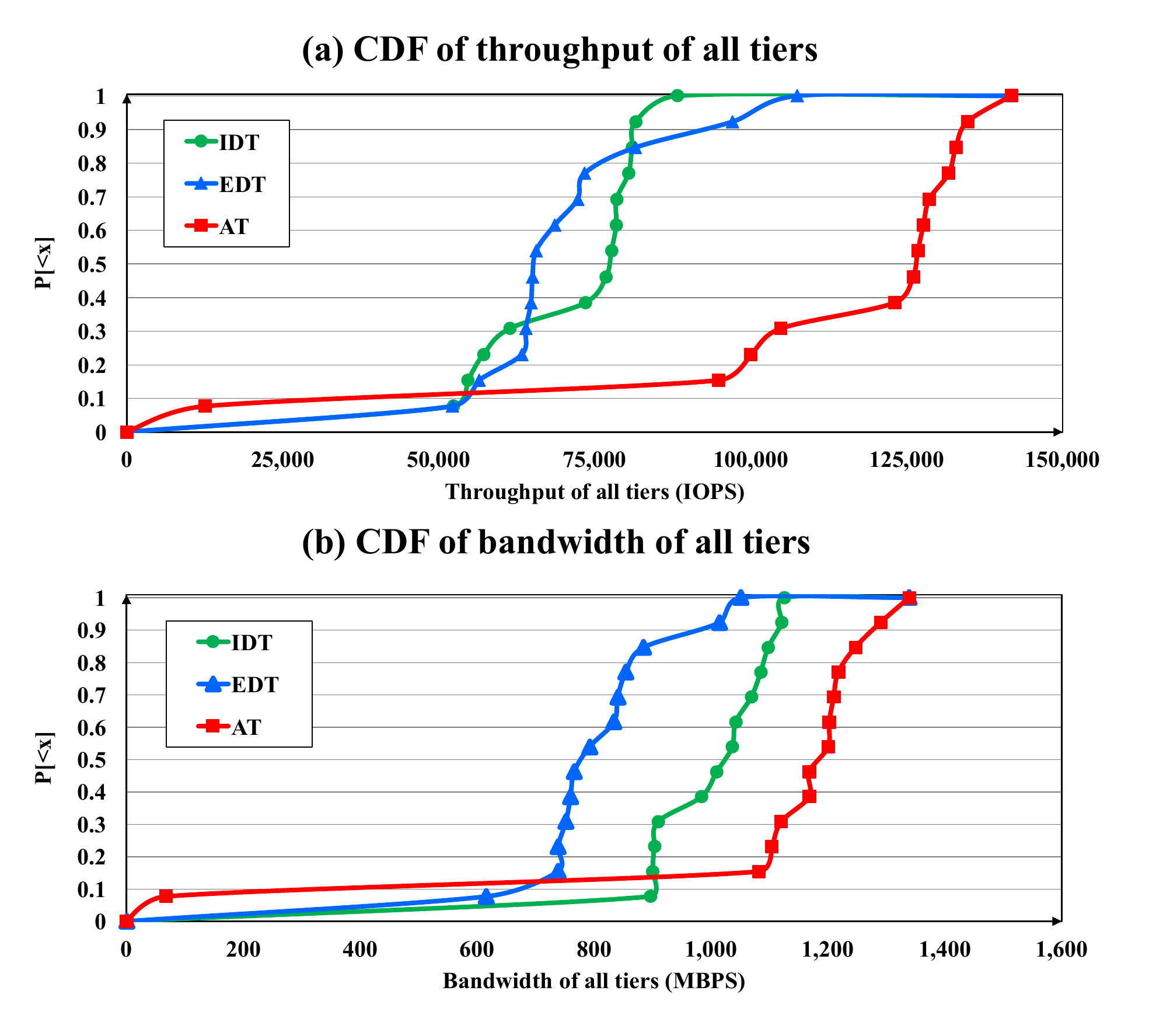}
\caption{\small CDF of throughput and bandwidth of all tiers.}
\label{FIG:REC-CDF}
\end{figure}

\begin{figure*}[ht]
\centering
\includegraphics[width=0.84\textwidth]{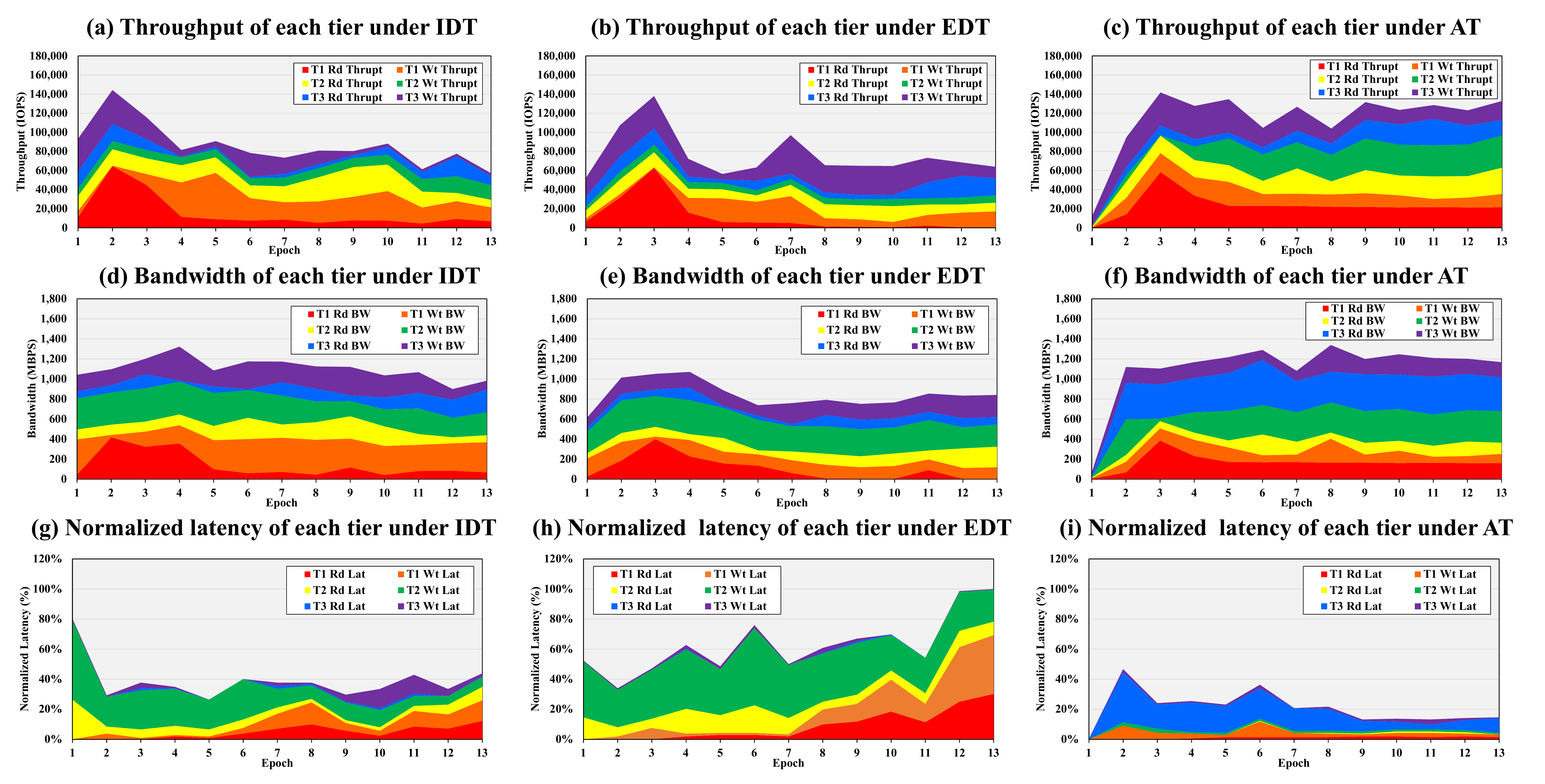}
\caption{\small Runtime changes of throughput, bandwidth, and latency of each tier under different algorithms.}
\label{FIG:REC-RUNTIME}
\end{figure*}


\begin{figure}[ht]
\centering
\includegraphics[width=0.34\textwidth]{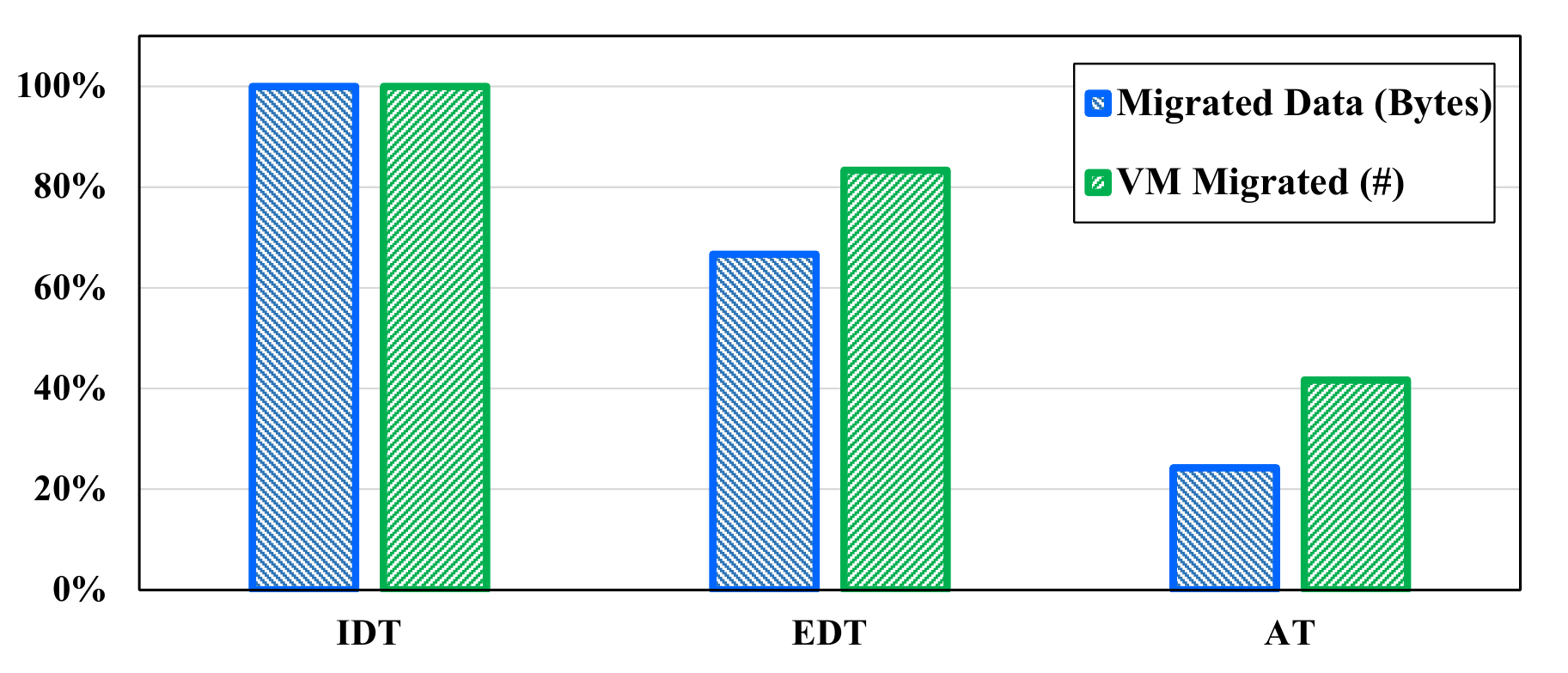}
\caption{\small Normalized migration cost results.}
\label{FIG:REC-COST}
\end{figure}

\vspace{-1mm}
\subsection{Evaluation Methodology}
\label{SUBSEC:METH}
\vspace{-1mm}

\subsubsection{Implementation Details}
We build AutoTiering on VMware ESXi hypervisor 6.0.0~\cite{esxi}. 
Table~\ref{TAB:EV-SPEC1} summarizes the server configuration of our implementation. 
Table~\ref{TAB:EV-SPEC2} further shows the specs of each tier (each tier has multiple SSDs).
We set the specialty matrix such that tier~1 is good for throughput and bandwidth performance,
tier~2 is the secondary performance tier but with larger capacity, and tier~3 is the capacity tier to replace HDDs.
%

\small
\begin{table}[ht]
	\center
	\caption{Host server configuration.}	
	\label{TAB:EV-SPEC1}
	\begin{tabular}{|c|c|}
		\hline
		\textbf{Component} & \textbf{Specs} \\ \hline 
		Host Server & HPE ProLiant DL380 G9  \\ \hline
		Host Processor &   Intel Xeon CPU E5-2360 v3 \\ \hline
		Host Processor Speed & 2.40GHz \\ \hline
    	Host Processor Cores & 12 Cores \\ \hline
		Host Memory Capacity & 64GB DIMM DDR4   \\ \hline
		Host Memory Data Rate & 2133 MHz \\ \hline
		Host Hypervisor & VMware ESXi 6.0.0 \\ \hline
	\end{tabular}	
	\end{table}
\normalsize

\subsubsection{Workloads}
To evaluate performance under different algorithms, we use IOMeter~\cite{iometer} and FIO~\cite{FIO} to generate I/O workloads to represent real world use cases. 
Table~\ref{TAB:TRACE} shows some statistical analysis of 14 used workloads.
Each VM has multiple VMDKs with different sizes, such as system disk, datastore disk, and etc.
%

\small
\begin{table}[ht]
	\center
	\caption{Multi-tier flash drivers configuration.}	
	\label{TAB:EV-SPEC2}
\begin{tabular}{|p{3.5mm}|p{21mm}|p{7mm}|p{4.5mm}|p{3.9mm}|p{4.4mm}|p{3.9mm}|p{10mm}|}
\hline
\multirow{2}{*}{\bf{Tier}}  & 	\multirow{2}{*}{\textbf{Model}} 	& \multirow{2}{*}{\textbf{Protcl.}} 		& \multicolumn{2}{l|}{\textbf{IOPS}}    & \multicolumn{2}{l|}{\textbf{MBPS}}  & {\bf{PerDisk}}    \\\cline{4-7}
    &      	&   		& {\bf{R}}   & {\bf{W}}     & {\bf{R}}  & {\bf{W}} & {\bf{Size(GB)}}       \\\hline
1 & Samsung PM953 & NVMe    & 240K & 19K & 1000 & 870 &   480      \\ \hline
2 & Samsung PM1633& SAS     & 200K & 37K & 1400 & 930 &   960       \\ \hline
3 & Samsung PM863 & SATA    & 99K  & 18K & 540  & 480 &   960     \\ \hline
	\end{tabular}	
	\end{table}
\normalsize


\vspace{-3mm}
\subsubsection{Comparison Candidates}
We compare AutoTiering ({\em{AT}}) with two other solutions~\cite{guerra2011cost}: (1) {\em{IDT}}: IOPS Dynamic Tiering, implements dynamic configuration and placement using a greedy IOPS-only criteria where higher IOPS extents move to higher IOPS tiers; and (2) {\em{EDT}}:  Extent-based Dynamic Tiering, updates VM-tier assignment for every epoch, based on both VM capacity and IOPS requirements.
To fully utilize the high speed all-flash datacenter, we slightly modified IDT and EDT to support {\em per-VMDK-based} operation.

\begin{table}[h]
\caption{Resource demands of selected workloads.}
\vspace{-1mm} 
\center
\begin{tabular}{|p{7mm}|p{15mm}|p{30mm}|c|c|}
\hline
\multirow{2}{*}{{\bf{Load}}} & \multirow{2}{*}{\bf{Workload}}& \multirow{2}{*}{{\bf{Represented Scenarios}}}	& {\bf{Thrupt.}}   & {\bf{BW.}}  \\
&   &     		& {\bf{(IOPS)}}    & {\bf{(BPS)}}                \\\hline
\multirow{5}{*}{Heavy}  & BasicVerify  &  SQL database server  & 95.5K   & 373M       \\ \cline{2-5}
 & SSDSteady   & System development  & 116K   & 453M         \\ \cline{2-5}
 & Zipf IOs        &   Web apps  &1942K  & 7585M       \\ \cline{2-5} 
 & AsyncRead     & Read intensive apps  &88.3K   &  345M   \\ \cline{2-5}
 & AsyncWrite    &  Write intensive apps  &6.65K  & 25M       \\ \hline
\multirow{5}{*}{Middle} & Flow   & Big data frameworks  &19.2K & 150M   \\ \cline{2-5}
 & IOmeter  & File server  &47K   & 205M       \\ \cline{2-5}
 & JESD   &  High endurance apps  &18.3K   & 136M      \\ \cline{2-5}
 & LatencyProfile  & Cloud system manager & 39.6K & 155M    \\ \cline{2-5}
 & SSDTest       &Hardware development  & 47K   & 205M       \\ \hline
\multirow{4}{*}{Light} & RandZone      & Multi-user database  &7.75K   & 30.3M      \\ \cline{2-5}
 & SurfaceScan    & Enterprise backup server  & 6.98K  & 436M       \\ \cline{2-5}
 & SyncRead      & Read intensive sync apps  &   6.65K  & 25M       \\ \cline{2-5}
 & SyncWrite     & Metadata sync server  &  4   & 16K       \\ \hline
\end{tabular}\label{TAB:TRACE}
\end{table}

\vspace{-1.5mm}
\subsection{Study on Throughput, Bandwidth and Latency Changes}
\label{SUBSEC:OVERALL}
\vspace{-1mm}
Fig.~\ref{FIG:REC-AVG} illustrates the average throughput, bandwidth, and normalized latency of all tiers over time for both read ({\em Rd}) and write ({\em Wt}) I/Os.
AutoTiering achieves up to 44.74\% and 38.78\% higher IOPS than IDT and EDT.
Similar results can be obtained for bandwidth and latency as shown in Figs.~\ref{FIG:REC-AVG}(b) and (c).
Fig.~\ref{FIG:REC-AVGTIER} depicts per-tier results to further show the performance improvement brought by AutoTiering.
We observe that AutoTiering performs the best in terms of (both read and write) throughputs, bandwidths and latencies on tier~1, which is because the specialty matrix sets tier~1 to optimize performance-sensitive workloads.
On the other hand, we also see that AutoTiering sometimes achieves lower throughput and bandwidth in tier~2 and 3 compared with IDT and EDT.
This is because IDT is IOPS-only algorithm, which migrates high-IOPS-demand (especially write I/O) workloads to tier~1, such that the write IOPS is optimized.
Similarly, EDT considers both IOPS and capacity, and thus has slightly better write IOPS compared to AutoTiering in the capacity tier~3.
It is worth to mention that AutoTiering achieves the lowest latencies in all cases except write latency in tier~2 (as shown in Fig.~\ref{FIG:REC-AVGTIER}(c) 5th column), because AutoTiering migrates many VMDKs that have large average I/O size (high write bandwidth), and thus, as a tradeoff, the latency is increased.   
%
Moreover, Fig.~\ref{FIG:REC-CDF} depicts the distribution of total throughput and bandwidth of all tiers for different algorithms.
From Fig.~\ref{FIG:REC-CDF}(a), we observe that under AutoTiering (red curve), majority of I/Os has more than 100K IOPS and even half of them have more than 125K IOPS. In contrast, 90\% of I/Os are less than 100K IOPS (blue curve) under IDT, and almost all I/Os from IDT are less than 100K (green curve).
Similarly, from Fig.~\ref{FIG:REC-CDF}(b), we can see that the majority (around 90\%) of IDT and EDT I/Os  are less than 1,200~MBPS, while more than half of AutoTiering's I/Os can achieve larger than 1,200~MBPS bandwidth.

\vspace{-2mm}
\subsection{Study on Runtime Distribution of Resource Utilization }
\label{SUBSEC:RUNTIME}
\vspace{-1mm}

Fig.~\ref{FIG:REC-RUNTIME} shows runtime changes of throughput, bandwidth and latency  distribution across tiers over time.
We observe that the areas in (c) and (f) are larger than those in (a)-(b), and (d)-(f), respectively.
Area in (i) is also much smaller than those in (g)-(h).
This verifies our observations in Sec.~\ref{SUBSEC:OVERALL} that AutoTiering achieves better throughput and bandwidth performance than IDT and EDT.
We also observe in (a) to (f) that area of each tier in AutoTiering is ``thicker'' than that in IDT and EDT (after AutoTiering's warming up periods from 0-3 epochs).
This indicates that AutoTiering can better utilize throughput and bandwidth resources of each tier by proper and less migrations.
In (i), we see that the majority of the latency of AutoTiering is located in tier~3 (the write latency ``{\em T3 Rd Lat}''), which is because that tier~3 is regarded as the capacity tier to replace HDD.
As a result, AutoTiering migrates read-intensive VMs with large VMDKs to leverage tier~3, and leaves tiers~1 and 2 for other write-intensive workloads.

\vspace{-3mm}
\subsection{Study on Migration Overhead}
\label{SUBSEC:MIGCOST}
\vspace{-1mm}
To investigate the migration overhead of three algorithms, we show the normalized temporal migration cost results in Fig.~\ref{FIG:REC-COST}.
The blue bars show the normalized total migrated data size, and the green bars show the normalized number of VMs that are migrated (multiple migrations on a single VM is counted as 1).
The former is to reflect the ``working volume size'' and the latter is to reflect the ``working set size''.
AutoTiering performs best among the three, since it migrates less data and interrupts less VMs, which saves lots of system resources.
In summary, AutoTiering chooses the best tier for each VM for better performance and prevents unnecessary migrations due to I/O spikes, ascribed to its comprehensive decision which is based on a more accurate performance estimation method.
\vspace{-2mm}
\section{Conclusion}
\label{SEC:CON}
\vspace{-1mm}

We present a novel data placement manager ``AutoTiering'' to optimize the virtual machine performance by allocating and migrating them across multiple SSD tiers in the all-flash datacenter.
AutoTiering is based on an optimization framework to provide the global best migration and allocation solution over runtime.
We further proposed an approximation algorithm to  solve the problem in a polynomial time, which considers both historical and predicted performance factors, and estimated migrating cost.
%
Experimental results show that AutoTiering can significantly improve system performance.
%

\vspace{-2mm}
\bibliographystyle{IEEEtran}
\bibliography{07_Reference}
\end{document}